**Studies of laser stimulated photodetachment from nanoparticles for particle charge measurements**

Y.A. Ussenov[1,*], M. N. Shneider[2], S. Yatom[1] and Y. Raitses[1]

[1]*Princeton Plasma Physics Laboratory, Princeton, NJ 08540*

[2]*Department of Mechanical and Aerospace Engineering,*

*Princeton University, Princeton, NJ 08544*

**e-mail: yussenov@pppl.gov*

## Abstract

Determining nanoparticle charge is more challenging than that for microparticles due to change in the particle size during the synthesis substantial plasma property variations, and difficulties in visualizing individual particles, rendering conventional microparticle charge diagnostics ineffective in dusty plasma. In this work, we utilized laser-stimulated photodetachment (LSPD) to deduce the mean charge of nanoparticles. Nanoparticles were grown in an $Ar/C_2H_2$ mixture using a capacitively coupled RF discharge and the LSPD induced changes in the electron current monitored by a cylindrical Langmuir probe. LSPD signals were obtained and analyzed across different dust growth phases. The prolonged decay of electron current pulses was attributed to the presence of residual negative ions, caused by the effective electrostatic trapping of these ions and the potential post - LSPD re-formation of new ones. The particle charge was estimated by combining the laser-stimulated photodetachment signal from the probe with the dust density obtained from laser-light extinction using the measured nanoparticle size distribution. For a nanoparticles size range of approximately *100 - 250 nm* and mean diameter of $d_p \sim 154.37$ *nm*, the effective mean charge was estimated to be $\langle Q_d \rangle \approx 37$ elementary charge units. The measured charge values are lower than those predicted by orbital motion limited (OML) theory, which may be attributed to significant electron depletion in the nanodusty plasma. LSPD results in $Ar/C_2H_2$ nano-dusty plasma confirm the applicability of this method for estimating individual nanoparticle charges. However, it has also been demonstrated that electron detachment from residual background negative ions can influence the detachment current decay and must be carefully considered.

## 1. Introduction

Dusty plasma, also known as "complex" plasma, is a type of plasma that contains small charged particles of solid or liquid matter, in addition to the usual ions, molecules, electrons, and neutral atoms found in regular plasma.[1] These dust particles can range in size from nanometers to micrometers and can acquire a significant charge as a result of incoming fluxes of ions and electrons.

Dusty plasmas are found in astrophysical environments like interstellar space,[2] comets, and planetary rings.[3] They are also subject of lab scale research to explore complex properties like self-organization, wave propagation, and crystal formation.[4] Dusty plasma occurs in magnetically confinement fusion devices (e.g. tokamaks and stellarators) due to the erosion of the walls under hot plasma flux, affecting performance, cooling, and fuel retention.[5] Therefore, dust powder in fusion devices has historically been considered an undesirable source of contamination.[6] In semiconductor plasma processing tools, dust particles typically act as contaminants that can cause defects on wafers, affecting device quality.[7] Generated through plasma-surface interactions, material erosion, and gas-phase nucleation, these particles can degrade device performance, making their control essential for high-quality semiconductor fabrication. For instance, in EUV lithography machines, dust particles reduce light generation efficiency and can damage optical components.[8] Formed through plasma-surface interactions, these particles absorb or scatter EUV radiation, degrade optics, and lower the precision of lithography process.

Dust nanoparticles can also be intentionally grown in plasma by the plasma-enhanced nucleation and growth,[9] where gas-phase materials condense into solid particles due to complex interactions in the plasma environment. This is a promising approach for generating functional nanoparticles[10] with the desired size and density for applications including medicine,[11] catalysis, quantum sensing and optoelectronics. It is anticipated that the "dusty plasma" synthesis offers better control of particle dispersity and avoids the use of wet chemicals, making it a cleaner, greener method.

Dust particles in plasma become charged by flux of electrons and ions, leading to electrostatic forces that influence plasma behavior. This charge plays a key role in nanoparticle growth and transport, altering electric fields, especially in the sheath region, and affecting plasma dynamics. The particle charge can be calculated by considering the balance between electron and ion fluxes to the particle, as well as the plasma properties such as temperature and density of charged components. For weakly collisional dusty plasmas with $\lambda_{in} \gg \lambda_D$, where $\lambda_{in}$ – ion-neutral collision mean free path, $\lambda_D$ – Debye length, the charging process is typically described using the orbital motion limited (OML) theory, which gives an estimate of the particle charge as a function of plasma parameters.[12] The implication of the OML theory for charge diagnostics is straightforward due to the known size of injecting microparticles and relatively non-sophisticated diagnostics of required plasma parameters. The accessibility of monitoring of the single microparticle trajectory or dynamics by the laser light scattering and video imaging allows to do particle charge diagnostics with so-called resonance methods.[13]

Unlike microparticle charge diagnostics, the determination of the charge of nanoparticles is more complex due to several reasons. These include the significant change in nanoparticle size during the growth, the strong correlation between plasma properties[14] and nanoparticle formation process,[15] low scattering properties, and the limitations of visualization methods. These factors prevent direct detection of nanoparticle dynamics or trajectories. Therefore, applying conventional methods for diagnosing microparticle charge to nanoparticles in dusty plasma faces certain challenges.

There are several alternative methods which are already applied for the diagnostics of grain charge in the nanodusty plasmas. The analysis of the so-called dust density or dust plasma acoustic waves in the nanoparticle cloud first was suggested as a tool to extract the particle size[16] and later was applied to define the dust charge and ion density in the amorphous hydrogenated carbon (a:C-H) nanodusty plasmas.[17,18] The experimentally measured dust density wave parameters linked with the results of the hybrid fluid-kinetic model and the quantities like dust charge, Havnes parameter[19], electron density, ion temperature, and electric field were estimated. In another approach, the IR absorption-based technique[20] was utilized to measure the $SiO_2$ nanoparticle charges confined in the RF plasmas. The shift in the IR absorption is attributed to the charging state of the nanoparticles. This shift, combined with the calculated shift from experimental data, was used to deduce the particle charge.[21,22] Finally, the individual dust particle charge was estimated in the $Ar/SiH_4$ dusty plasmas using the difference in ion and electron density.[23] The electron and ion densities were measured by the shielded fast-scanning

Langmuir probe and capacitive probe respectively. The size of the particles analyzed by ex-situ transmission electron microscopy. Using the proposed analytical model,[24] the density and charge of the particles were estimated.

Another promising approach to measure the negative charge of nanoparticles in dusty plasma is the laser stimulated photo detachment of surface electrons[25] (LSPD), which is studied in this paper. The LSPD was initially developed and demonstrated by M. Bacal and colleagues for the study of negative ions.[26] Subsequently, it has been widely used for measurements in low-temperature plasmas[27,28] and ion beam sources.[29] This technique has also been applied to investigate the formation and charging dynamics of negatively charged dust particles.[30] The LSPD involves using a laser to detach electrons from negatively charged particles (dust grains or negative ion) through the photodetachment process, which allows for the measurement of the particle charge or charge density. Detached electrons create a photoelectron current in the plasma, which can be detected using electrical probes or microwave-based methods. This current is proportional to the number of electrons detached. Therefore, for dust particles, it is related to the particle charge. If the particles density monitored with other in-situ methods (e.g., laser induced incandescence[31] (LII), the charge per single particle can be estimated. Unlike many other techniques that infer charge from indirect measurements (e.g., plasma parameters, cloud dynamics), LSPD allows for a relatively straightforward measurement of the charge on negatively charged particles with minimum perturbation of the plasma. The latter is because the laser power is not high enough to evaporate the particles and because the increase of the electron density induced by LSPD from particles is much less than the equilibrium plasma density.

The LSPD was applied to the dusty plasma in Ar/$SiH_4$ mixture to detect the initial nuclei formation process at early stages of the discharge ignition.[32] As the threshold energy depends on the particle size, the obtained photo current results were interpreted involving both photo detachment and photoemission mechanisms. Recently, the particle size effect on the LSPD also studied theoretically, predicting that the charging of nanoparticles in plasma leads to the appearance of an additional electric field.[33] This causes a change in the potential barrier at the particle boundary and, consequently, a change in the effective work function due to the Schottky effect. The LSPD also has been recently applied in the Ar/$SiH_4$[30] and in hexamethyldisiloxane (HMDSO)/Ar[34] plasma. LSPD-induced changes in the electron density were measured with the microwave resonant cavity spectroscopy. The particle charge density and charge per particle along with nanoparticle recharging time were estimated. The latter was used for monitoring the size of nanoparticles[35,36] assuming the electron temperature, ion density and temperature. Compared with other available approaches, LSPD provides a relatively direct and time-resolved probe of electron release from nanoparticles, i.e.,charging dynamics, without a need in complex optical arrangements or a complete set of local plasma parameters.

Herein, we study the LSPD in carbonaceous nanodusty plasma grown in an Ar/$C_2H_2$ mixture, aiming to estimate the charge of individual nanoparticles in reactive hydrocarbon plasmas. The Ar/$C_2H_2$ discharge was chosen as a well-established model dust-forming plasma[15,18], in which carbonaceous nanoparticles reproducibly form and the underlying plasma chemistry and particle-growth dynamics have been widely studied, while avoiding the handling complexity of highly reactive precursors such as silane. Changes in the plasma density induced by LSPD were monitored using a single Langmuir probe. The nanoparticle density measured with laser light extinction while a mean nanoparticle size was obtained for the ex-situ evaluation of synthesized nanoparticles using scanning electron microscopy (SEM). The LSPD was conducted for different particle growth phases to examine the impact of the electron photodetachment from background negative ions. The formation and sources of negative ions after $C_2H_2$ flow termination, along with their role in the LSPD signal, were analyzed. The paper is organized as follows: Section 2 details the experimental procedure, Section 3 presents and discusses the results, and Section 4 provides conclusions and outlines directions for future work.

## 2. Experimental setup and procedure

A schematic of the experimental setup is shown in Figure 1. Dust nanoparticles were grown in a capacitively coupled radio frequency (13.56 MHz) discharge (RF-CCP) at constant 10W RF power, driven by a R301 generator coupled with an ATS3 matching box (both from Seren IPS). Two parallel stainless-steel electrodes, 10 cm in diameter, were placed in the central region of the vacuum chamber with a discharge gap of ~38.0 mm. The bottom electrode was powered, while the top electrode was grounded. Prior to the experiments, the chamber was pumped down to a base pressure of ~$5x10^{-6}$ Torr using a turbo-molecular pump backed with a mechanical pump. The plasma was generated in an argon-acetylene gas mixture produced by a 50 sccm Ar flow (Linde, Ultra High Purity), controlled by a needle valve coupled with a mass flow meter (Tylan, FM-360), and a 4 sccm $C_2H_2$ flow (Linde Gas), controlled by a mass flow controller (Bruker, UHP series). The gas flow was introduced from one of the gas injection ports on the side of the chamber while the vacuum pump port located in the front of this port. Prior to the plasma generation, a gas mixture pressure of 480 mTorr was established and maintained in the chamber by adjusting the mechanical throttle valve on the vacuum pump line. During the particle growth, the pressure increased up to ~530 mTorr due to the decomposition of $C_2H_2$ molecules. The gas temperature was assumed to be equal to room temperature. At a low RF power of 10 W, gas heating is negligible because energy transfer from energetic electrons to heavy neutrals is highly inefficient. Furthermore, any minor thermal energy is rapidly dissipated to the chamber walls, maintaining the gas temperature effectively at room temperature. Unless stated otherwise, the mentioned parameters were kept the same for all experiments.

Nanoparticles were grown in plasma through the dissociation of $C_2H_2$ molecules, followed by the nucleation and agglomeration of the dissociation byproducts[10,38]. They are charged negatively in the plasma and therefore, confined within the plasma volume between the two electrodes due to a balance of electrostatic forces, gravity, and neutral and ion drag. However, after reaching the certain particle size and mass they may fall or escape from the plasma volume, potentially forming thin films composed of dust grains covering the bottom electrode or discharge chamber walls.

The nanoparticle growth process was monitored by measuring a DC self-bias voltage ($V_{DC}$) at the RF powered electrode and the laser light extinction (LLE) signal. The $V_{DC}$ signal was extracted from the analog control port of the ATS3 matching box (with 1:200 ratio) and recorded by the oscilloscope (Tektronix TDS 2014, 1 MHz, 1 Gs/s). For the LLE, a red He-Ne laser pointer (5 mW, 632.8 nm) was directed through the nanodust cloud close to the Langmuir probe tip, and the extinction signal was measured by a Si photo detector (Thorlabs, DET36A), coupled with a neutral density filter, and monitored by an oscilloscope (Tektronix TDS 2014, 1 MHz, 1 Gs/s). The synthesized nanoparticles were collected on silicon wafer coupons (10 mm × 10 mm) fixed on an aluminum substrate holder. During the initial growth cycles, the substrate holder was left floating at a negative potential, so negatively charged dust particles were electrostatically repelled and deposition was strongly suppressed. During the LSPD measurements, the holder was biased to +25 V, switching it into a collection regime. Thus, the SEM size distribution mainly reflects particles collected during the same time window as the LSPD diagnostics rather than the full growth history. The substrate was placed at the periphery of the electrodes along the gas-flow direction, so the collected particles are expected to originate from the discharge volume contributing to the line-of-sight LLE and LSPD signals. Thin films formed on silicon substrate were further analyzed using scanning electron microscope (SEM). To reduce bias from sticking, we did not treat large agglomerates as single particles and only measured primary particles individually when their boundaries were clearly resolved, including in chain-like deposits.

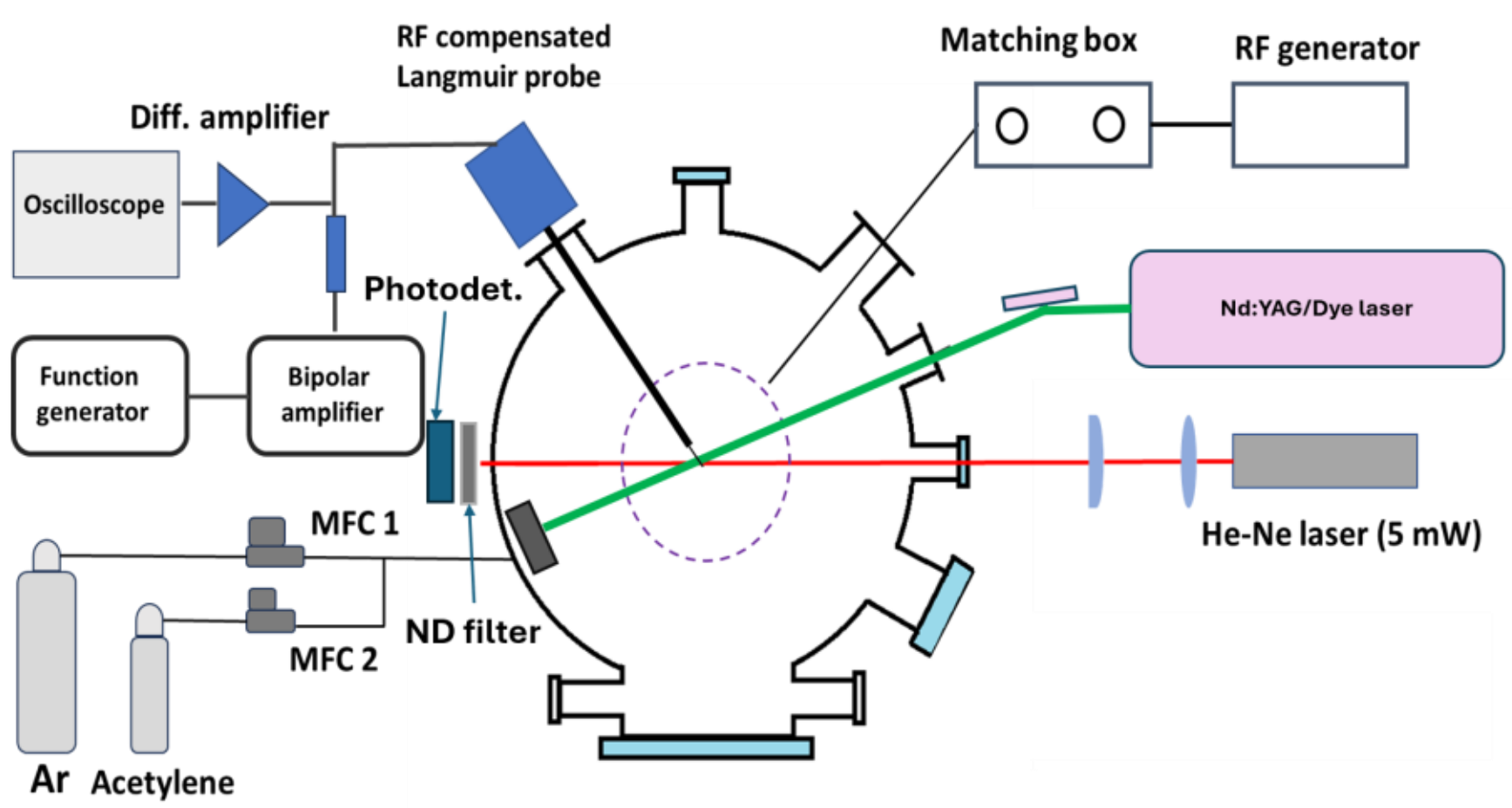

Figure 1. The schematic of the experimental setup for LSPD from dust nanoparticles in RF-CCP discharge.

For the laser stimulated photodetachment, the third harmonic of the Nd:YAG laser beam (Continium, Surelite III) at 355 nm (10 Hz, ~9 ns pulse width) was used. The laser beam energy varied between 2-75 mJ. To conduct time resolved measurements of the electron density using the RF compensated single Langmuir probe (10 mm length, 114.3 μm radius) biased by the bipolar amplifier (Kepco BOP 72-5M) and signal generator (DDS - FY6800). The laser and probe bias signals are synchronized by the external trigger output of the Nd:YAG laser control unit and BNC 7010 digital delay generator.  The current signals were recorded with Siglent SDS 2354X HD (350 MHZ, 2 Gs/s) oscilloscope through the 5.1 kΩ shunt resistor and Teledyne LeCroy AP031 (1:10, 25 MHz) differential probe. To eliminate the probe tip contamination in reactive hydrocarbon plasma the probe tip kept outside the plasma region during the main growth phase and biased -20V with respect to ground to repel negatively charged dust particles. Then, prior to starting the LSPD the probe was positioned in the main plasma region. A +50V positive bias with ~1 ms pulse duration was applied to the probe to keep it in the electron saturation current regime. It is in this regime, the changes of the probe current due to LSPD from the dust particles were recorded.

In typical dusty plasma experiments, the spatial distribution of nanoparticles within the discharge gap can be non-uniform[37] due to the combined effects of ion drag, electrostatic confinement, gravity, and gas flow. This may result in the formation of a dust free region at the center of the discharge during the particle growth cycle. To avoid this region, the probe tip was positioned slightly toward the plasma periphery so that the measurements were performed in a dust-containing region of the discharge. Using optical hardware, the laser beam was directed to pass through the probe tip fully covering the tip length. The surplus of electrons generated by the LSPD resulted in short time changes (up to ~300 μs) of the probe current with respect to its steady state value. This current change was recorded by the oscilloscope. After the single LSPD event, the probe bias was returned to -20 V to avoid dust particle collection by the tip and to keep a continuous flux of positive ions to the surface. Between the dust growth cycles, the probe tip was periodically cleaned in a pure Ar plasma by combination of ion bombardment and electron heating regime at the -80V and +80V probe bias, respectively.

## 3. Results and Discussion

### *3.1 Particle growth and experimental phases.*

To accurately estimate the particle charge from the measurements, it is essential to minimize electron photodetachment from negative ions and maintain a constant dust particle density during the LSPD. Therefore, the entire experiment duration was divided into several phases, which were monitored by the $V_{DC}$ and LLE. Typical results are shown in Figure 2(a) and 2(b), respectively. The time evolution of these parameters indicates the presence of three distinct phases[34]: (i) the particle growth phase under a constant flow of $C_2H_2$, (ii) the stabilization phase after turning off the $C_2H_2$ to eliminate acetylene molecules, dissociation byproducts and negative ions, and (iii) the LSPD measurement phase. First, the discharge chamber is filled with the $Ar/C_2H_2$ mixture, then the RF discharge plasma is initiated. The $V_{DC}$ is always negative, and during the first few milliseconds, it sharply increases (in absolute values) to ~45-60 V. The corresponding sharp peak is marked as “Plasma ON” in Figure 2. This is due to a high flux of electrons reaching the cathode following the rapid ionization of Ar atoms and $C_2H_2$ molecules. Afterward, the density of $C_2H_2$ dissociation byproducts reaches a critical level, promoting polymerization which involves negative ions electrostatically trapped in the positive plasma volume. According to model by De Bleecker et. al., [38] the dominant negative ions, which are effectively trapped in the plasma volume and serve as precursor molecules for the nucleation and polymerization growth of nanoparticles, are $C_2H^-$ (ethynyl), $H_2CC^-$ (vinylidene, formed by the reaction of $C_2H^-$ with $C_2H_2$ molecules), and higher-order molecular anions, such as $C_{2n}H_2^-$. Polymerization of the negative ions with molecules leads to the formation of initial nanoclusters. Agglomeration of these nanoclusters probably resulted in the formation of larger dust nanoparticles with radii of approximately ~5-10 nm.[39] The nanoparticles are negatively charged by electrons, resulting in significant electron loss to the surface of nanoparticles which causes a observable drop in $V_{DC}$ down to ~3-6 V (Figure 2(a)). Once the nanoparticles accumulate a stable negative charge, they can no longer agglomerate due to a Coulomb repulsion. However, they continue to grow through the "surface growth" mechanism, which involves the accretion of defragmentation byproducts of acetylene molecules in the plasma.[40] This growth is slower than the particle growth by agglomeration of nanoparticles. Overall, the dust particle growth period lasts approximately 200 seconds and before the $C_2H_2$ flow was shut down. During this time, the LLE signal gradually decreases over the first 100 seconds, confirming the active formation of nanoparticles by the agglomeration process and then, it stabilizes at a saturation level. After the nanoparticle growth phase, the stabilization phase began, during which the $C_2H_2$ flow was turned off while the Ar flow and pumping were maintained. Despite the termination of precursor feed into the chamber, a drop in both the $V_{DC}$ and LEE was observed, indicating further nanoparticle formation. This phenomenon can be attributed to the enhanced ionization of remaining $C_2H_2$ molecules, driven by the increased effective energy per molecule under constant RF discharge power. In Figure 2a, the starting point of this phase coincides with the point marked as "$C_2H_2$ OFF". This phase lasts another ~200 seconds. After the stabilization phase ended, the LSPD measurement phase was initiated. During this phase, the LLE signal changed by less than <5%, allowing us to assume that the particle density during the LSPD events and the electron density do not change.

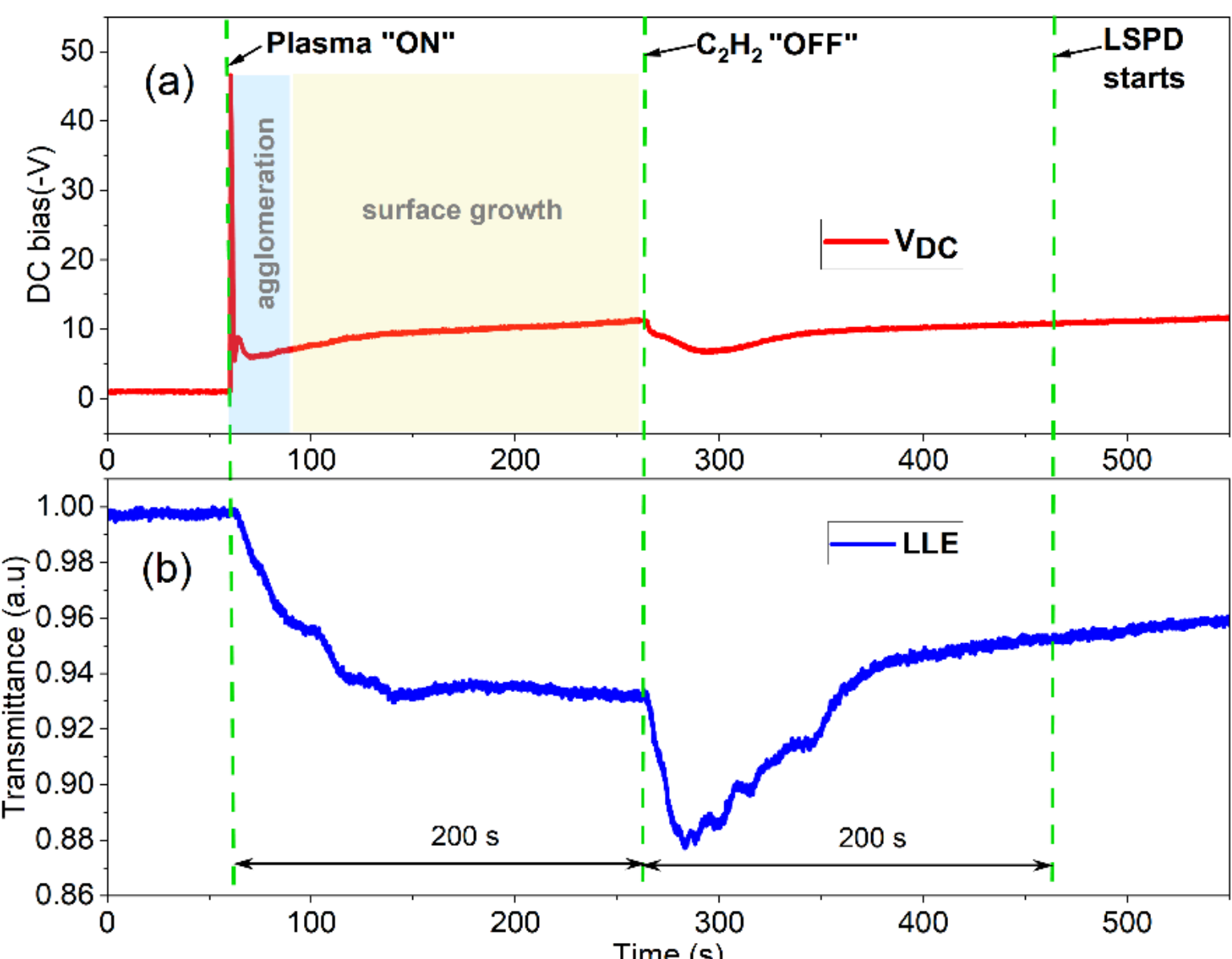

Figure 2. (a) Discharge self-bias voltage signal and (b) laser light extinction signal during the dust particle growth, stabilization, and measurement phases. The LLE signal varied by less than 5% during the LSPD phase, suggesting that particle density remained nearly constant.

### *3.2 Dust density*

Estimations of the nanoparticle density require information on the mean particle size (diameter). The SEM micrographs (Figure 3(a)) show that the particles are spherical with a "cauliflower-like"[40] surface structure. Size distribution analysis, performed using ImageJ software (Figure 3(b)), showed that a mean particle diameter $d_p = 154.37\ nm$, with an overall distribution range of ~100-250 $nm$.

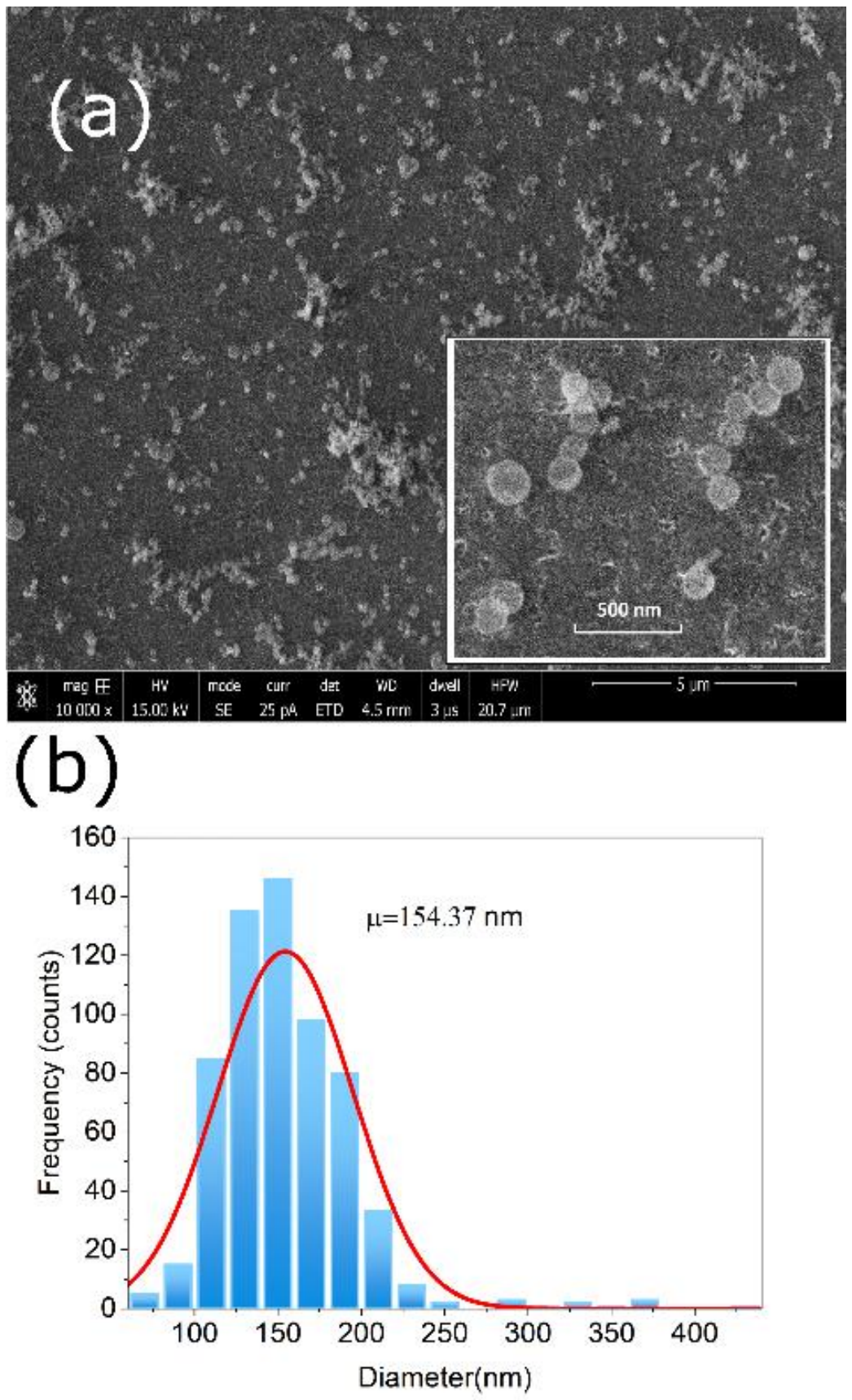

Figure 3. SEM micrographs of the collected dust nanoparticles (a) and the corresponding size distribution histogram (b). The inset in (a) shows a zoomed-in image of a set of individual nanograins, with the scale bars indicated in the images.

Using the measured mean diameter of nanoparticles and the light extinction of the laser passing through the dust cloud, we can estimate the dust particle density. According to the Beer-Lambert law[41], the attenuation of light passing through a medium filled with particles can be described by the optical depth, $\tau = -\ln\,(I/I_0)$, where $I_0$ and $I$ are the incident and transmitted laser light intensities, respectively. Then, for a monodisperse particle population, the particle density can be estimated using the [34,42] $n_d = \frac{\tau}{\pi r_d^2 Q_{ext} L}$, where $n_d$ is the dust density, $r_d$ - dust nanoparticle radii, $L$ - optical path length, $Q_{ext}$- nanoparticle extinction efficiency. Optical depth $\tau \approx 0.3$. $Q_{ext}$-is a dimensionless parameter that describes how well light attenuates due to the combination of scattering and absorption. It is determined by the particle's refractive index $N$, the wavelength of incoming laser light (632.8 nm), and the particle size from SEM analysis.

Since the SEM results in Fig. 3(b) show a broad particle size distribution rather than a monodisperse population, the extinction analysis was performed using a polydisperse Mie-scattering[43] approach. For each size bin $i$ of the measured histogram, the extinction efficiency $Q_{\text{ext}}(r_i)$ was calculated at $\lambda = 632.8$ nm using the refractive index $N = 1.93 + i0.043$ for amorphous hydrogenated carbon (a:C-H) nanoparticles[44], and the corresponding extinction cross section was determined as:

$$C_{\text{ext},i} = \pi r_i^2 Q_{\text{ext}}(r_i). \tag{1}$$

The histogram counts were then used to define the normalized number fraction $f_i = N_i / \sum_i N_i$, and the distribution-averaged extinction cross section per particle was obtained from:

$$\langle C_{\text{ext}} \rangle = \textstyle\sum_i f_i \; C_{\text{ext},i} = \sum_i f_i \; \pi r_i^2 Q_{\text{ext}}(r_i). \tag{2}$$

Accordingly, the particle density was estimated from:

$$n_d = \frac{\tau}{L\langle C_{\text{ext}} \rangle}. \tag{3}$$

Using the measured size histogram, the extinction cross section evaluated at the mean particle diameter ($d_p = 154.37$ nm) is $C_{\text{ext,mean}} = 5.81 \times 10^{-15}$ m$^2$ and the dust particle density is $n_d = 4.96 \times 10^7 \; cm^{-3}$. Whereas the distribution-averaged value is $\langle C_{\text{ext}} \rangle = 1.35 \times 10^{-14} m^2$. Thus, the polydisperse treatment gives an average extinction cross section larger by a factor of 2.32 than the monodisperse mean-diameter approximation. For the same measured optical depth and optical path length, this yields a dust particle density of $n_d = 2.14 \times 10^7 \text{cm}^{-3}$. These values are consistent with typical densities for dust nanoparticles grown using RF-CPP with hydrocarbon precursor gases[18] and used below for the estimation of the effective mean nanoparticle charge from the LSPD measurements.

*3.3 LSPD at different discharge conditions.*

To investigate the impact of negative ions formed through the dissociation of hydrocarbon molecules, LSPD measurements were conducted in a pure Ar plasma as well as in dusty plasma with and without an active $C_2H_2$ flow. Typical probe current signals for LSPD in different regimes are shown in Figure 4.

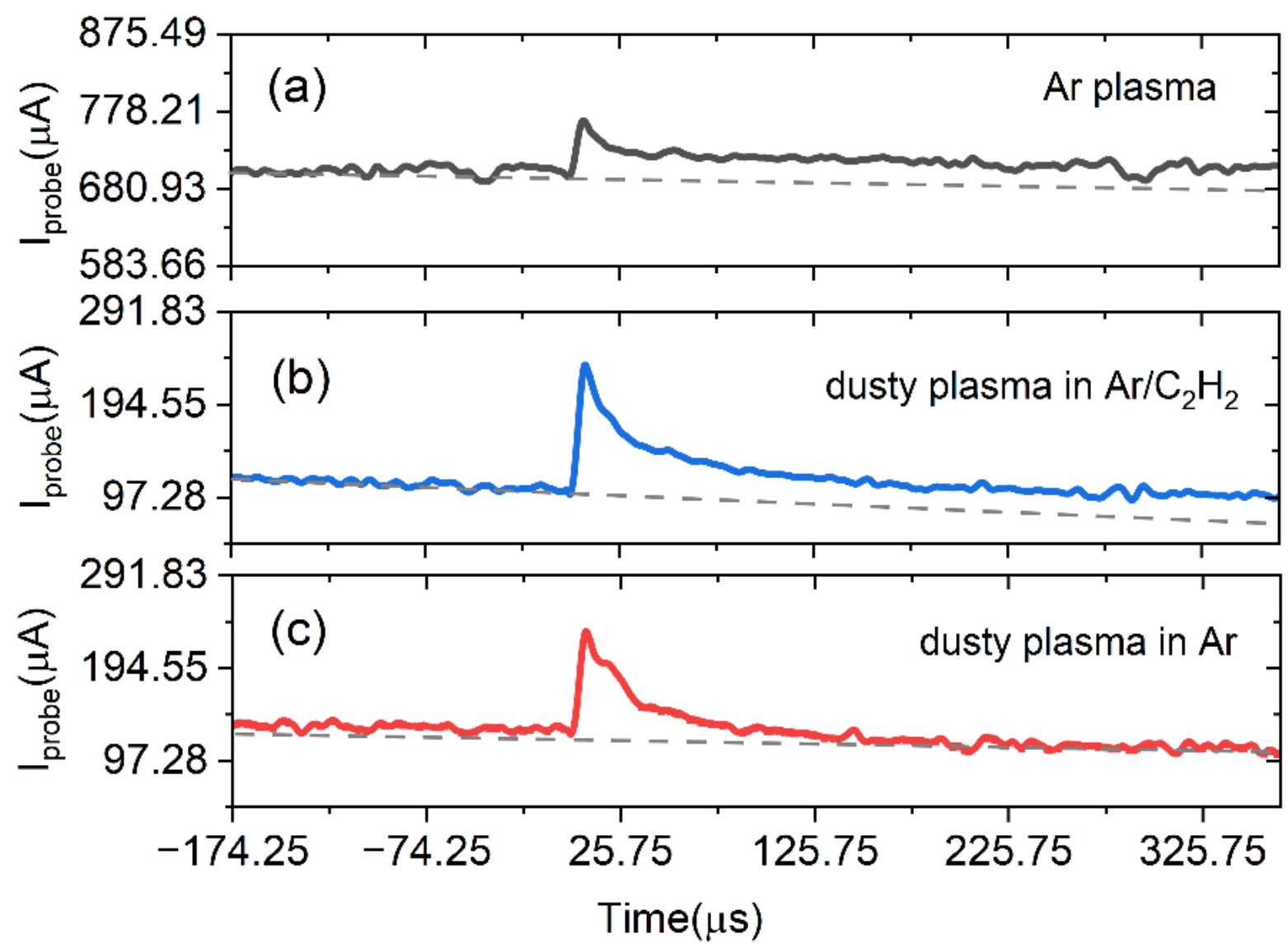

Figure 4. The typical Langmuir probe current in the electron saturation regime during LSPD events obtained in three conditions: pure Ar plasma, dusty plasma in an Ar/$C_2H_2$ mixture, and dusty plasma in Ar after stopping the $C_2H_2$ flow.

In a pure Ar plasma, the LSPD induces a ~10% increase of the probe current, from 705 μA to 775 μA (Figure 4a). This weak LSPD signal is attributed to the electron photodetachment from negative ions. The generation of these ions can be attributed to the desorption of water molecules and hydrocarbon contaminants from the electrode surface and chamber wall.[45] Note that simple calculations (detailed in Appendix A) suggest that both photoelectric and thermal emissions from the probe tip resulting from laser irradiation are negligible.

In a separate experiment, plasma was generated in an Ar/$C_2H_2$ mixture, and dust nanoparticles were grown for 200 seconds. Afterward, the probe was inserted into the dust cloud region, and LSPD was performed. The results of LSPD (Figure 4b) in dusty plasma under the constant flow of $C_2H_2$ and Ar show a much larger increase of the probe electron current from the $I_{pr} = 145.9\ \mu A$ to the $I_{pr} = 244\ \mu A$ (Figure 4b) than for pure Ar plasma. It is worth to mention that in Ar/$C_2H_2$ dusty plasma (Figure 4b), the baseline probe electron current is generally lower compared to pure Ar plasma (Figure 4a). This is due to the reduction of electron density caused by plasma wall losses on dust particles. The full decay time of the LSPD signal is prolonged and does not return to the base value within ~375 μs.

Following this experiment a new dust growth cycle was started and LSPD was performed in dusty plasma with a pure Ar (Figure 4c). In this case, the $C_2H_2$ flow was shut off after dust growth, and the chamber was pumped for 200 seconds under a constant 50 sccm Ar flow. The LLE signal was monitored during the pumping period. After reaching a stable particle density, the Langmuir probe was inserted into the nanodust cloud, and LSPD was performed. Under such conditions, the LSPD signals exhibit a peak value similar to that of the dusty plasma case in the $C_2H_2$/Ar mixture (Figure 4c). However, the decay time is shorter, with the signal fully decaying after approximately 130 µs. The LSPD signal decay time reflects the decay of surplus electron density in the plasma and is determined by the re-attachment of electrons to the dust particles (i.e. particle recharging) and residual negative ions. A typical recharging timescale for dust nanoparticles depends on their size (cross section) and is relatively short, ~5-6 µs.[34] The prolonged decay time of the probe electron current after the LSPD laser shot shown in Figure 4 may be attributed to the reattachment of electrons to electronegative molecules of the background residual $C_2H_2$ gas leading to the (re-) formation of negative ions.

To examine the LSPD in the Ar/$C_2H_2$ mixture with negative ions, experiments were conducted during the early plasma initiation period, i.e. prior to dust grain formation. Figure 5(a) shows a typical temporal variation of $V_{DC}$ during the first several seconds of the plasma generation in $C_2H_2$/Ar gas mixture. During this phase, $C_2H_2$ molecules start to dissociate forming byproducts, including negative ions, and singly charged nanoclusters with size <5 nm. [46] At the same time, the formation of larger dust nanoparticles driven by coagulation and agglomeration did not start. Although the $V_{DC}$ signal rapidly decreases after ~500 ms, its nearly flat shape up to 300 ms indicates minimal loss of plasma electrons. The latter suggests the absence of nanoparticles. At this early growth phase, LSPD signal can be attributed to the detachment of electrons from negative ions, such as $C_2H^-$,[47] which are primarily formed through the dissociative attachment of electrons to acetylene molecules. The LSPD signals taken at both 100 ms and 200 ms after plasma ignition are shown in Figure 5 (b). In addition, the corresponding time steps are indicated as a vertical dashed line in the insert of Figure 5 (a). These timepoints are selected as an example; in the interval prior to 300 ms, when $V_{DC}$ remains stable, indicating that nanoparticle formation has not started yet. Both LSPD signals demonstrate an increase of the electron current measured with the probe, up to ~34 µA. This current corresponds to the $\Delta N_e \sim 6.45 \times 10^7\ cm^{-3}$. As will be shown in Figure 6, this is an order of magnitude lower compared to the case with nano dusty plasma, where $\Delta N_e \approx 7.8 \times 10^8 cm^{-3}$. This result suggests that unlike the case with nanoparticles, there is no multiple electron stripping and thus, the source of detachment is mainly negative ions. Although the LSPD signal amplitude is relatively weak, its decay time of approximately ~150-200 μs is comparable to that observed for photodetachment from nanoparticles in dust-containing plasmas in the $Ar/C_2H_2$ mixture and in pure Ar dusty plasmas. This similarity suggests that, in all cases, the prolonged decay time is indicative of the reattachment of electrons to negative ions.

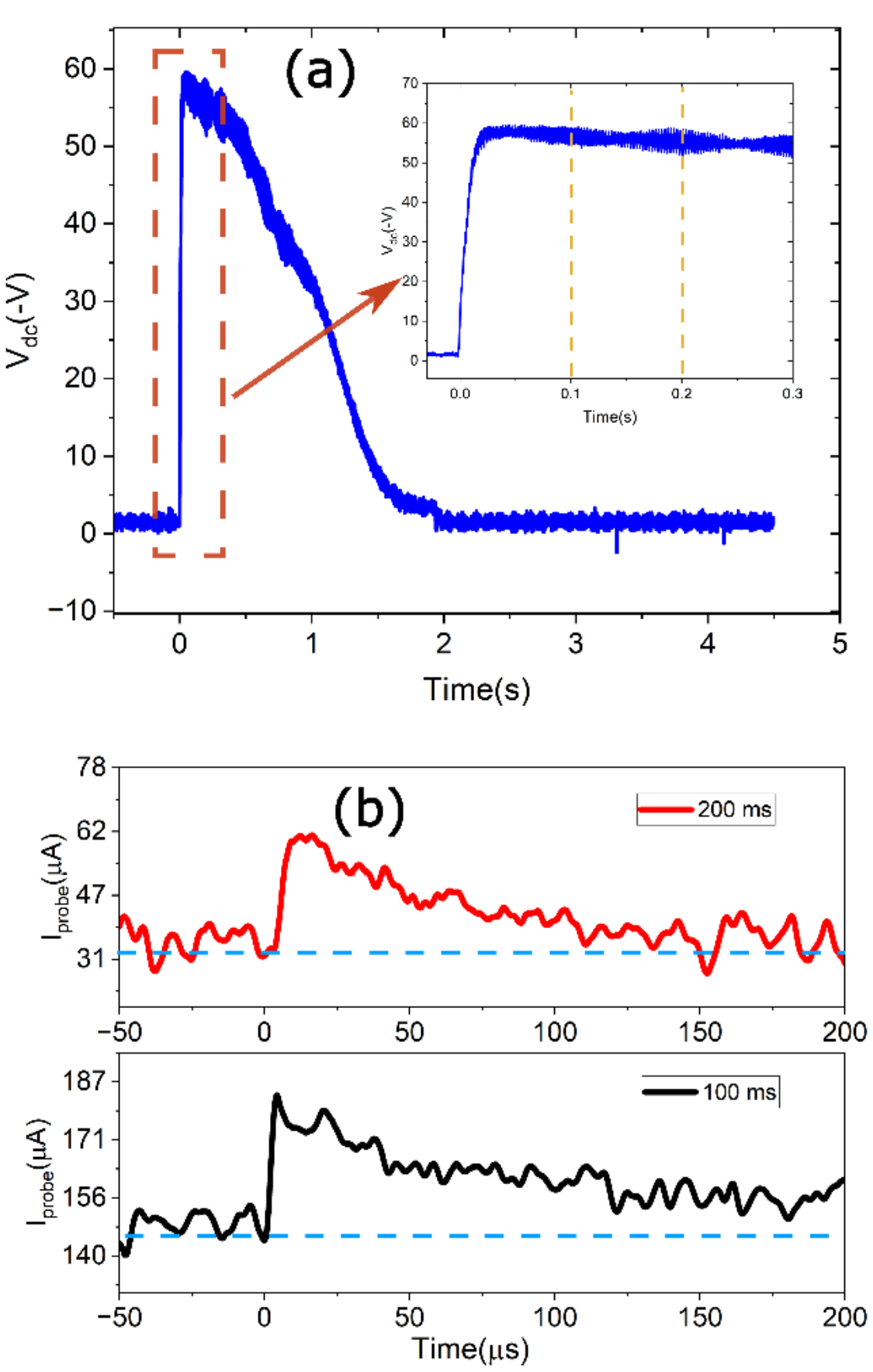

Figure 5. Discharge self-bias voltage $V_{DC}$ during the early phase of nanoparticle formation (a), with the inset depicting its temporal evolution over ~300 ms. The time steps at which the LSPD is performed are marked by vertically dashed yellow lines. The corresponding LSPD signals at the 100 ms and 200 ms time steps shown in (b).

Similarly prolonged LSPD signal decay times were also observed during the early growth phase (before nanoparticle formation) in $Ar/C_2H_2$[46] and $SiH_4/Ar$[30] plasmas and attributed to the re-formation of negative ions after laser pulse.

### *3.4 The source of residual negative ions without $C_2H_2$ flow*

The termination of $C_2H_2$ and pumping under the continuous Ar flow would significantly reduce the concentration of monomer $C_2H_2$ molecules - and, consequently, the negative ions. However, the assumption that neutral drag induced by Ar neutrals would efficiently pump out all the negative ions may not hold, as they are electrostatically trapped in the plasma volume.[48] In this scenario, the remaining negative ions may be lost through the formation of new nanoparticle clusters. This process is driven by their polymerization with residual $C_2H_2$ molecules during the short period after stopping the acetylene flow. Hence, the primary photodetachment reactions from the negative ions for the LSPD in our case (with laser photon energy $h\nu$ = 3.5 eV) are identified along with their corresponding electron affinities $E_a$[49,50]:

$$H_2CC^- + h\nu \rightarrow C_2H_2 + e^- \quad (E_a = 0.49\ eV)$$

$$C_2H^- + h\nu \rightarrow C_2H\cdot + e^- \quad (E_a = 2.956\ eV)$$

Another finding that supports the presence of negative ions in dusty Ar plasma is the growth of nanoparticles without active $C_2H_2$ flow. This growth occurs due to sputtering of films deposited on the cathode surface, resulting in the formation of new nanoparticles. The growth of these nanoparticles was confirmed through a separate set of experiments, where pure Ar plasma was initiated after several growth cycles in a $C_2H_2/Ar$ mixture. More details are provided in Appendix A. Additionally, recent observations indicate that already-formed large nanoparticles can continue to grow[51] in pure Ar dusty plasma through a process analogous to "Ostwald ripening".[52] Furthermore, it is well established that a-C:H nanoparticles undergo continuous etching[17] in pure Ar plasma. In both cases, the formation and trapping of negative ions in the plasma volume play a key role in the growth of nanoparticles.

In summary, a relatively prolonged decay of the LSPD signal in a plasma result from the reattachment of photodetached electrons to negative ions and molecules in nanodusty Ar plasmas. These are due to (i) effective trapping of residual negative ions in the gas flow, (ii) byproducts of sputtering from the a-C:H thin layer on the cathode and wall surface, and (iii) etching and evaporation of very small a:C-H dust nanoparticles. Direct monitoring of negative ions under LSPD experimental conditions would require highly sensitive negative ion mass spectrometry, which poses challenges due to the difficulty in extracting electrostatically trapped negative ions[53]. This aspect is outside the scope of this paper. Nevertheless, the contribution of the long-decay component in the LSPD signal is relatively minor. This is because the photodetachment from negative ions leads to the striping of single electrons, while detachment from nanoparticles results in the release of multiple electrons.

*3.5 Nanoparticle charge.*

The saturation of the LSPD signal is expected to manifest that all charges electrons were detached from the particle surface. To test this hypothesis LSPD measurements were conducted at varying laser beam energies. Figure 6 shows the LSPD signals for $E_{laser}$ from ~1.8 mJ to 75.6 mJ. The density of detached electrons $\Delta Ne$, as a function of laser energy, is shown in the inset of Figure 6. After 45.15 mJ, the LSPD signals start to reach a saturation value. The probe electron current at +50V bias is converted to electron density using the formula for electron saturation current for Maxwellian electron energy distribution:

$$I_{es} = \frac{1}{4} e n_e A_p \sqrt{\frac{8kT_e}{\pi m_e}}, \quad (4)$$

where $I_{es}$- electron saturation current, $n_e$- electron density, $A_p$- surface area of the probe tip, $m_e$ – electron mass, $kT_e$- electron temperature. The above expression is consistent with the OML theory approach, and generally satisfies the $r_p < \lambda_D < \lambda_{mfp}$ criteria, where the probe radii $r_p = 0.11\ mm$, and Debye length $\lambda_D \approx 0.81$ mm and electron - neutral collision mean free path is $\lambda_{emfp} \approx 1.25\ mm$ for our experimental conditions.

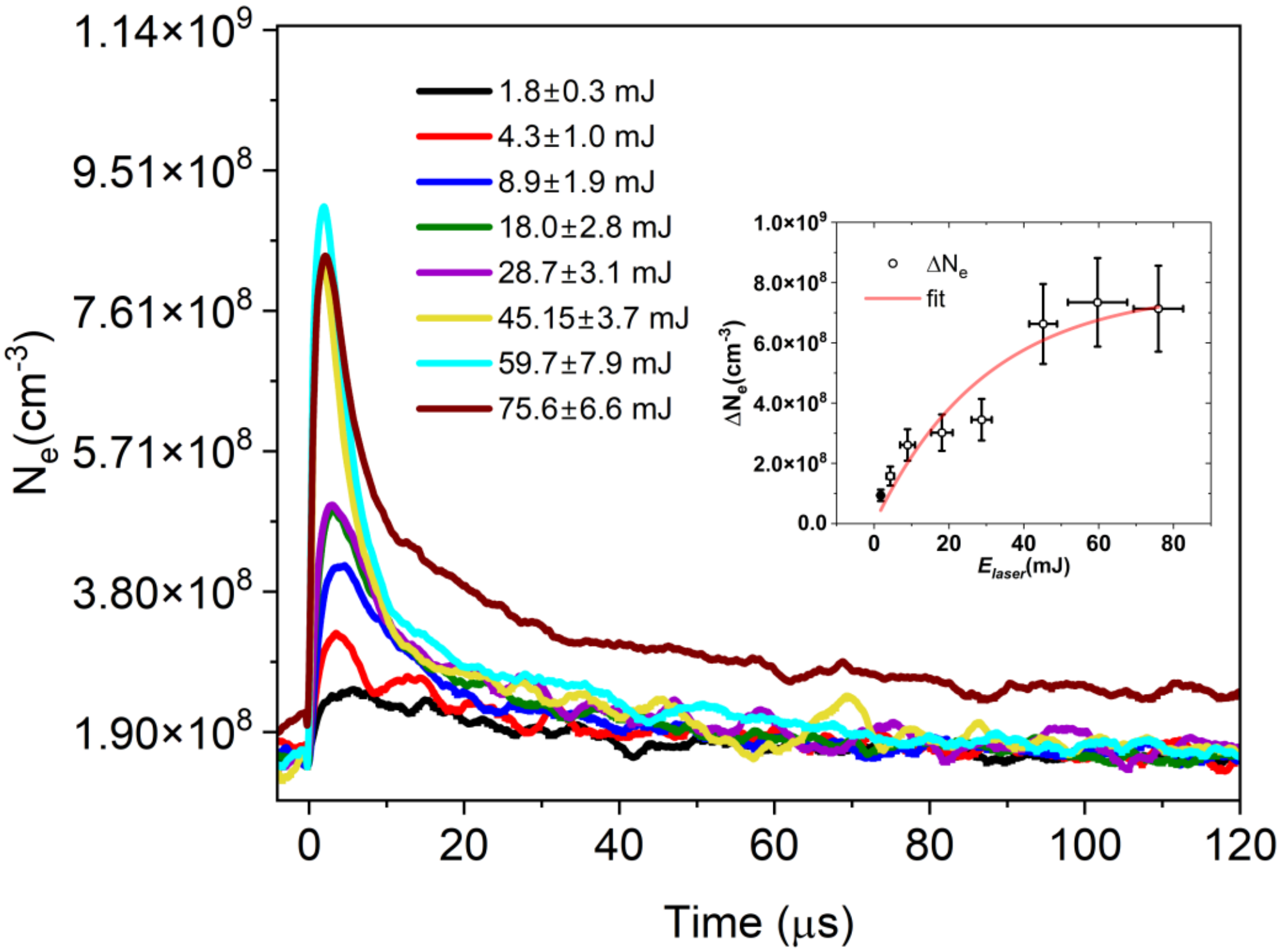

Figure 6. The change in the plasma electron density measured during the LSPD for different $E_{laser}$ values. The peak density of detached electrons, fitted using expression 3, are shown in the inset figure.

The fraction of the photodetached electrons[54] at certain $E_{laser}$ can be described as $\alpha \cong \frac{\Delta N_e}{N_e} = (1 - \exp(-\frac{\sigma_{pd} E_{laser}}{hvS}))$, where $\Delta N_e$ density of detached electrons, $N_e$ – the overall density of electrons residing on the dust particle surfaces, $\sigma_{pd}$ -photodetachment cross section, $E_{laser}$- laser beam energy, $hv$ – photon energy, $S$ – laser beam area. To estimate the $\sigma_{pd}$ from the saturation curve of the detached electron density at different $E_{laser}$, the fitting of the detached electron density vs the laser energy was performed following the expression:[34]

$$\Delta N_e = N_e(1 - \exp(-\frac{\sigma_{pd} E_{laser}}{hvS})), \quad (5)$$

The fitting results gives a photodetachment cross section $\sigma_{pd} = 1.2 \times 10^{-20}\ m^2$, at the saturation value of $N_e = 7.8 \times 10^8\ cm^{-3}$. The estimated cross-section value is smaller than the values obtained for SiOx nanodusty plasmas.[34] This difference can be attributed to the difference in material properties of hydrocarbon (a-C:H) particles, growth conditions and lower LSPD photon energy (3.5 eV compared to 4.66 eV).

The change of the plasma density during the LSPD at saturated $E_{laser}$ = 59.7±7.9 mJ is shown in Figure 7, along with the double exponential decay fitting. The density of the detached electrons, estimated from the full height of the LSPD signal relative to the steady-state plasma density, is equal to $\Delta N_e$ = 7.8 x $10^8$ $cm^{-3}$. The first exponential decay time constant is $\tau_1 = 3.67\,\mu s$ which is expected and in a reasonable agreement with the recharging time of the nanoparticles in the electron depleted dusty plasmas. The second exponential time decay constant is $\tau_2 = 45.1\,\mu s$ is attributed to the slow decay of the LSPD signal due to the recharging of the residual background negative ions.

To estimate the nanoparticle charge from the saturated LSPD signal, the surplus photodetached electron density $\Delta N_e$ is related to the dust density obtained from the LLE measurements. For the present conditions, the saturated LSPD signal gives $\Delta N_e$ = 7.8 x $10^8$ $cm^{-3}$. Using the LLE inversion evaluated at the mean particle diameter from the SEM analysis, $d_p$ = *154.37 nm*, the corresponding dust density is $n_d = 4.96 \times 10^7\ cm^{-3}$ , which gives a mean charge of $Q_d \approx \frac{\Delta Ne}{n_d}$ ≈*16e*. This monodisperse estimate is retained here as a reference value because it reflects the result obtained when the extinction is evaluated only at the mean particle diameter. However, the SEM analysis in Fig. 3 indicates a broad particle size distribution, and therefore the charge estimate can be refined by using the polydisperse extinction treatment introduced in Sec. 3.2. With this approach, the inferred dust density becomes $n_d = 2.14 \times 10^7 \mathrm{cm}^{-3}$, which yields an effective mean nanoparticle charge of $\langle Q_d \rangle \approx 37e$. This value should be interpreted as an effective mean charge of the particle population sampled by the combined LLE/LSPD measurement geometry, rather than as the charge of a single particle with diameter 154.37 nm.

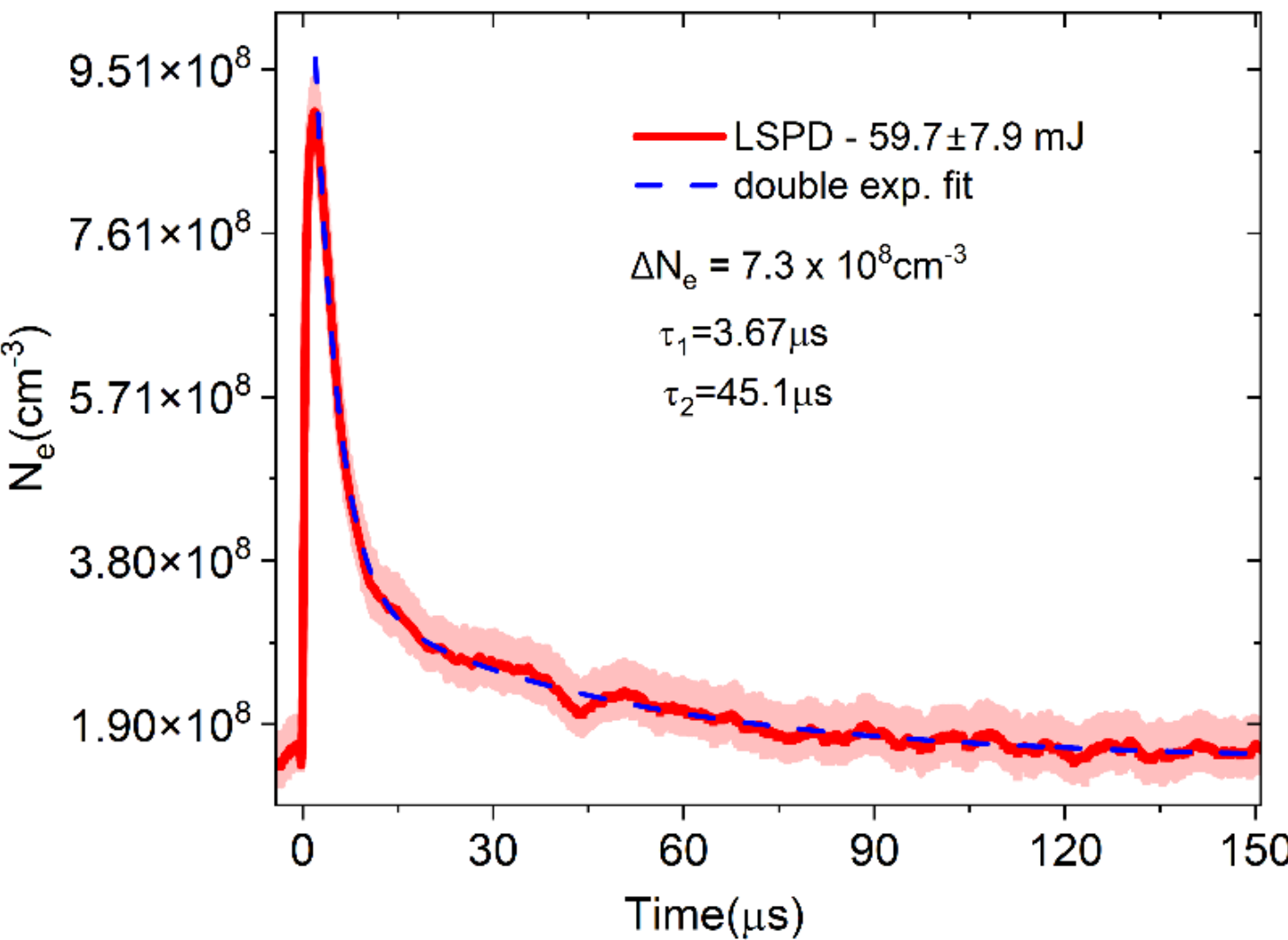

Figure 7. The change in plasma density detected by the probe during the LSPD for $E_{laser}$=59.7±7.9 mJ, presented alongside the corresponding double exponential fitting curves.

Since the LSPD is a line integrated measurement along the laser beam, we can expect changes in the size distribution of nanoparticles along this line in the discharge volume, which is also confirmed by ex-situ SEM analysis (Figure 3). The particle density $n_d$ was re-estimated for the maximum and minimum particle diameters obtained from the size distribution histogram (Figure 3) using Eq (3). For particles with diameters $d_p$ = *100 nm* and $d_p$ = *250 nm,* estimated $n_d = 4.81 \times 10^8\ cm^{-3}$ and $n_d = 1.75 \times 10^6\ cm^{-3}$ respectively. For these particle densities, the corresponding nanoparticle charge lies in the range of $Q_d \approx$ *5 - 130e.* These limiting cases are not intended to represent realistic monodisperse dust clouds, but rather to show the uncertainty introduced when a broad size distribution is reduced to a single representative diameter. Accordingly, in the discussion below, the polydisperse result, $\langle Q_d \rangle \approx$ 37e, is taken as the best estimate for the effective mean nanoparticle charge under the present experimental conditions, while the values obtained from the mean, minimum, and maximum diameters are retained as useful reference points for the sensitivity of the estimate to the assumed particle size.

The estimated charge values were compared with selected data from other studies[17,18,20] (more details in Table 1 in Appendix). Taking into account the variations in particle growth conditions and the methods used for estimating the dust charge, the results of the present experiments are found to be in reasonable agreement with previous results.

We also compared the experimentally obtained charge with theoretical values estimated using the OML approach for a single nanoparticle in dusty plasma. According to the capacitance model the particle charge approximately follows the practical expression[55] $Q_{d,OML} \approx 1400 r_d T_e$ , where $r_d$ is dust radius in $\mu m$ and $T_e$ is electron temperature in eV. For our experimental conditions $Q_{d,OML} \approx 630$e. This is significantly higher than the experimentally obtained value $\langle Q_d \rangle \approx 37e$ from the LSPD measurements. This discrepancy is consistent with strong electron depletion in RF discharge plasmas with high nanodust density. This electron depletion phenomenon is known as the Havnes effect[19] and is common in nanoparticle or submicron-sized complex plasmas, where most electrons in the plasma volume reside on the surfaces of particles. The Havnes parameter $P$ can be expressed using the following formula:[19,42]

$$P = 4\pi\varepsilon_0 \frac{d_p}{2} \frac{k_B T_e}{e^2} \frac{n_d}{n_i} \quad (4)$$

where $d_p$- dust mean diameter, $k_B$-Boltzmann constant, $k_B T_e$ – electron temperature, $n_d$ -dust density and $n_i$ -density of ions. Assuming the ion density in the range[56,57] of $n_i \sim 10^9$-$10^{10}$ cm$^{-3}$, in our case the $P \cong 1 - 7$. This aligns well with the results obtained in the $C_2H_2$/Ar nanodusty plasmas. A P>1 indicates an electron depletion regime and a significant change in the background plasma properties due to the presence of dust grains. Under these conditions, the low nanoparticle charge observed in our experiments is expected, particularly given the smaller dust particle size detected compared to similar studies.

## Conclusions

The extended decay of Langmuir probe electron current signals following LSPD events in Ar/$C_2H_2$ nanodusty plasmas highlights the significant role of residual negative ions. This was linked to residual negative ions remaining in the plasma even after the gas supply was stopped. Their persistence might be caused by the factors such as electrostatic trapping of existing ions and the generation of new ones. New ions may form as molecules enter the plasma through sputtering from electrodes or vacuum discharge chamber walls, along with material released from nanodust grains during etching processes.

The dust particle density was determined from laser light extinction using the SEM-derived size histogram and a polydisperse Mie-scattering model. For particles with a mean diameter of about ~154.37 nm, we computed the extinction cross section for each size bin and obtained a distribution-averaged value, yielding a dust density of ~$2.1 \times 10^7$ $\mathrm{cm}^{-3}$. Combining this density with the saturated LSPD electron yield of ~$7.8 \times 10^8$ $\mathrm{cm}^{-3}$ gives an effective mean nanoparticle charge of ~$37e$ elementary charges. These obtained inferred charges are significantly lower than simple single-particle OML predictions $Q_{d,OML} \approx 630\mathrm{e}$ for isolated grains of similar size and plasma parameters but are consistent with strong electron depletion expected in RF discharges with high nanodust loading. Since both LLE and LSPD are line-of-sight diagnostics, the reported dust density and charge represent geometry-averaged estimates and remain subject to uncertainty from possible spatial nonuniformity of the dust cloud.

Overall, our results show that LSPD, when combined with probe measurements and polydisperse laser-extinction analysis, provides a practical and minimally invasive diagnostic in the bulk plasma for estimating an effective mean nanoparticle charge in reactive nanodusty plasmas. At the same time, they highlight that residual negative ions and line-of-sight averaging of both dust density and size distribution introduce systematic uncertainties that must be explicitly taken into account. The role of negative ions could be significant for measurements in flow tube reactors, where eliminating negative ions by trapping the dust cloud and continuous pumping under noble gas flow may be challenging.

Future work will focus on investigating how the inferred charge and negative-ion contribution depend on particle size distribution, growth phase, and discharge conditions, and on developing complementary diagnostics to better resolve the spatial structure of both the dust and negative-ion populations. More accurate nanoparticle optical properties, especially the complex refractive index and dielectric function at the diagnostic wavelengths, would improve the reliability of Mie-based extinction calculations and the inferred particle parameters. Better constraints on electron affinity, work function, and plasma-modified surface states would also strengthen the interpretation of LSPD signals.

**Appendix**

The photoemission from the tungsten probe tip during the laser shots are unlikely, as the photon energy $h\nu$=3.5 eV is much lower than the work function of the tungsten W~4.5-5.0 eV. To assess the possibility of thermal emission, the maximum temperature rise $\Delta T_{max}$ of the tungsten probe tip surface after laser pulse exposure was estimated by the[58] $\Delta T_{max} \approx 0.22(1 -$ $R)E_{laser}/(d^2\sqrt{kC_p\rho\tau})$, where $\Delta T_{max}$ is the maximum change in temperature (at ~ 0.55 $\tau$), $R$ - reflectance at given wavelength, $E_{laser}$-laser beam energy, $d$- beam diameter, $k$ - heat conduction coefficient, $C_p$ - heat capacity, $\rho$ - mass density, $\tau$ - pulse duration. For highest $E_{laser}$~$80\ mJ$ the $\Delta T_{max}$ ≈76 $K$, meaning that the maximum surface temperature may rise up to ~376 $K$. This confirms the absence of thermal emission from the probe tip due to laser exposure.

Figure A1 shows the change in the $V_{DC}$ and LLE signal after plasma ignition in pure Ar gas, following the several cycles of nanoparticle growth in an Ar/$C_2H_2$ mixture. Similar to Figure 2, the $V_{DC}$ sharply increases after plasma ignition, which is indicated as "Plasma ON". It then gradually falls along with the LLE, confirming the formation of a dust nanoparticle cloud. Under these conditions, dust nanoparticles slowly form even without introducing $C_2H_2$ into the discharge chamber. The formation of nanoparticles is initiated by trapped negative ions and further enhanced by molecular fragments.[45] These ions and molecular fragments arise due to the sputtering of an amorphous hydrogenated carbon film and a thin nanoparticle-containing layer on the electrode surfaces, which formed during previous growth cycles in the Ar/$C_2H_2$ mixture. This sputter induced growth of dust particles in pure Ar plasma justifies the presence of negative ions in the discharge volume even after the stopping the $C_2H_2$ flow.

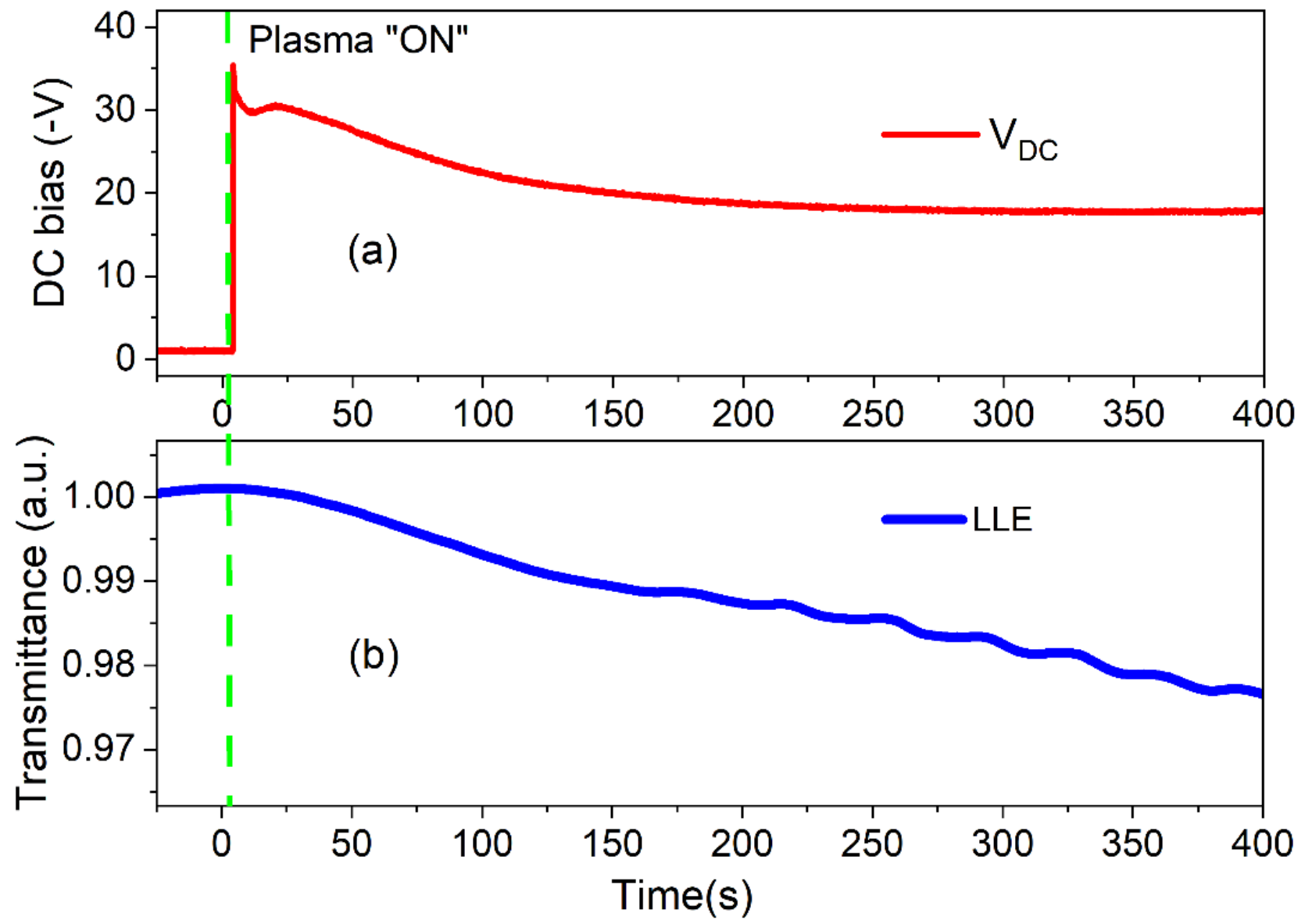

Figure A1. The change in $V_{DC}$ (a) and LLE (b) during the formation of dust particles in a pure Ar plasma after dust growth cycles in Ar/$C_2H_2$ mixture.

Table 1. The selected particle charge measurement results obtained for various nanodusty plasma conditions.

| Particles, radius | Method | Charge, e | References |
|---|---|---|---|
| a:C-H, 196 nm | Dust density waves | ~14-59 | B. Tadsen et.al., Phys. Plasmas 22, 113701 (2015)<br>https://doi.org/10.1063/1.4934927 |
| a:C-H, 120-160 nm | Dust density waves | ~10-40 | A. Peterson et.al., *Commun Phys* **5**, 308 (2022)<br>https://doi.org/10.1038/s42005-022-01060-5 |
| $SiO_2$, 40 nm | Infrared phonon resonance shift | ~168 | H. Kruger et.al., *Phys. Rev. E* 104, 045208 (2021)<br>https://doi.org/10.1103/PhysRevE.104.045208 |
| $SiO_x$, 12-75 nm | Ion-electron density difference | ~50-150 | N. Bilik et.al, *J. Phys. D: Appl. Phys.* 48 (2015) 105204<br>http://dx.doi.org/10.1088/0022-3727/48/10/105204 |
| $SiO_x$, ~5 nm | LSPD | ~25 | E. Stoffels et.al, *J. of Vac. Sc. & Tech.* A 14,556 (1996)<br>https://doi.org/10.1116/1.580144 |
| $SiO_xC_y$(HMDSO based), 140 nm | LSPD | ~273-2519 | T.J.A. Staps et.al., *J. Phys. D: Appl. Phys.* 55 (2022) 08LT01 |
| a:C-H, 77.18 nm | LSPD | ~37 | This work |

**Data Availability**

The data that support the findings of this study are available from the corresponding author upon reasonable request.

**Acknowledgements**

The authors acknowledge the US Department of Energy (DOE) for support of the work by the contract DE-AC02-09CH11466 and Mr. T. Bennet and Mr. A. Merzhevskiy for technical assistance. Y.U. acknowledges Prof. J. Beckers and Dr. T. Donders for the fruitful discussions during the $36^{th}$ ICPIG and $16^{th}$ Dusty Plasma Workshop.

**Conflict of interest**

The authors have no conflicts of interest to disclose.